\documentclass[twoside,12pt]{article}
\def\ifundefined#1{\expandafter\ifx\csname#1\endcsname\relax}
\makeatletter
\usepackage{theorem}

\newcommand{\reportenum}[3][ ]{\gdef\rep@rtenum{#2}
\gdef\rep@rteyear{#3}\gdef\wh@reappear{#1}}

\let\@ldmaketitle=\maketitle
\renewcommand{\maketitle}{{\def\newpage{}
{\scriptsize\parbox[t]{0.3\textwidth}{\noindent Reporte Interno
\# \rep@rtenum\\
Departamento de Matem\'aticas\\CINVESTAV del IPN\\Mexico City, \rep@rteyear}
\hfill
\parbox[t]{0.5\textwidth}{\wh@reappear}}
\@ldmaketitle}}

{\AtBeginDocument{
   \makeatletter
   \pagestyle{myheadings}
   \markboth{\hfill\ifundefined{@authorshort} \@author
                   \else \@authorshort
                   \fi\hfill}
            {\hfill\ifundefined{@titleshort} \@title
                   \else \@titleshort
                   \fi\hfill}
   \makeatother
 }
}

\ifundefined{thm}
     \theorembodyfont{\slshape}
        \newtheorem{thm}{Theorem}[section]
     
     \newtheorem{lem}[thm]{Lemma}
     \newtheorem{cor}[thm]{Corollary}

     \theorembodyfont{\upshape}

     \newtheorem{example}[thm]{Example}

     \newtheorem{rem}[thm]{\mdseries\scshape Remark}
\fi

\newenvironment{proof}[1][\proofname]{\par
  \normalfont
  \topsep6\p@\@plus6\p@ \trivlist
  \item[\hskip\labelsep\scshape
    #1{.}]\ignorespaces
}{%
  $\qed$\endtrivlist
}
\newcommand{\proofname}{Proof}
   {\@definecounter{equation}}{\@newctr {equation}[section]}

\newcommand{\epigraph}[3]{\par
\hfill\parbox{0.4\textwidth}{\footnotesize #1 \par \hfil #2 \it
#3}\par\medskip}
\providecommand{\dedicatory}[1]{}
\providecommand{\keywords}[1]{\begingroup \def \protect {\noexpand \protect
\noexpand }\xdef \@thefnmark { }\endgroup \@footnotetext{{\em Keywords and
phrases.\/} #1}}
\providecommand{\AMSMSC}[2]{\begingroup \def \protect {\noexpand \protect
\noexpand }\xdef \@thefnmark { }\endgroup \@footnotetext{{1991 \it
Mathematical Subject Classification.\/} Primary: #1; Secondary: #2.}}

\newcommand{\authorshort}[1]{\gdef\@authorshort{#1}}
\newcommand{\titleshort}[1]{\gdef\@titleshort{#1}}

\newcommand{\comment}[1]{}

\usepackage{amsfonts}
\newcommand{\algebra}[1]{\ensuremath{{\mathfrak #1}}}

\newcommand{\object}[1]{\ensuremath{\mathrm{#1}}\,}
\newcommand{\Space}[2]{\ensuremath{ {{\mathbb #1}^{#2}} }}

\newcommand{\FSpace}[2]{{\ensuremath{ #1_{#2} }}}

\newcommand{\such}{\ensuremath{\,|\,}}

\ifundefined{qed}
    \DeclareMathSymbol{\qed}{0}{AMSa}{"03}
\fi
\newcommand{\Cstar}{$C^{*}$}

\providecommand{\eqref}[1]{\textup{(\ref{#1})}}

\authorshort{Vladimir V. Kisil}
\titleshort{Relativistic Quantization}

\reportenum[]{179}{1995}

\begin{document}
\title{ Relativistic Quantization and\\
Improved Equation for \\a Free Relativistic Particle%
}
\author{Vladimir V. Kisil\thanks{On leave from the Odessa State
University.}\\
Departamento de Matem\'aticas,
CINVESTAV del I.P.N.,\\
Apartado Postal 14-740,
07000, M\'exico, D.F. M\'EXICO  \\
\normalsize fax: (525)-752-6412\ \
e-mail: \ttfamily vkisil@mvax1.red.cinvestav.mx}
\date{February 16, 1995}
\maketitle
{\em Dedicated to the memory of my grandfather M. M. Poggio
(1907--1994)}
\enlargethispage*{50pt}
\begin{abstract}
Usually the only difference between relativistic quantization and
standard one is that the Lagrangian of the system under consideration
should be Lorentz invariant.
The standard approaches are logically incomplete and produce
solutions with unpleasant properties: negative-energy, superluminal
propagation etc.

We propose a two-projections scheme of (special) relativistic
quantization. The first projection defines the quantization procedure
(e.g. the Berezin-Toeplitz quantization). The second
projection defines a casual structure of the relativistic system
(e.g. the operator of multiplication by the characteristic
function of the future cone).
The two-projections quantization introduces in a natural way the
existence of three types of relativistic particles
(with $0$, $\frac{1}{2}$, and $1$ spins).
\AMSMSC{81P10, 83A05}{81R30, 81S99, 81V45}
\end{abstract}
\newpage
\tableofcontents
\section{Introduction}
\epigraph{One should keep the need for a sound mathematical basis
dominating one's search for a new
theory.}{P.~A.~M.~Dirac~\cite{Dirac78}.}{}
The paper presents a new scheme of relativistic quantization.
Our approach is different from other ones (for example, the
Dirac construction). Usually the only difference between relativistic
quantization and  standard one is that  the Lagrangian of the system
under consideration should be Lorentz invariant. Quantization
procedures themselves are the same
for non-relativistic and relativistic cases. After solutions to the
Schr\"odinger equation are obtained, one should consider additional
restrictions: delete negative-energy solutions, forbid superluminal
propagation or introduce second quantization to describe ensembles of
bosons or fermions.

Another approach to relativistic quantization was used by
Dirac~\cite{BogShir80,Dirac78,Thaller92}  to
introduce an equation for a free relativistic electron. This approach
has the following features:
\begin{enumerate}

\item\label{it:deduction} Its deduction contradicts all usual
quantizations.

\item\label{it:spin} It gives a natural description of the spin of
an electron and its magnetic moment.

\item\label{it:negative} It involves the inconvenient occurrence of
states with negative energy with many unpleasant consequences (for
example, {\em Zitterbewegung\/}).

\end{enumerate}

One can suspect, that item~\ref{it:negative} is a corollary
of~\ref{it:deduction}, but it was unknown how one can obtain
item~\ref{it:spin} without the false prerequisites and
{\em negative\/} outcomes. We try to improve this situation by introduction a
two-projections scheme of relativistic quantization.

The first
projection defines quantization procedure (for example, the
Berezin-Toeplitz quantization). The second projection defines a casual
structure in relativistic system (for example, the operator of
multiplication by the characteristic function of the future cone in
the tangent space to space of events). We do {\em not\/} use in our
construction the Lorentz (or another relativistic) group. The main
idea of the presented quantization is an application of the causality
constraint before the derivation of motion equations.

For the given energy operator $H$ the two-projections quantization
gives us a family of algebras of observables with the Heisenberg
motion equation for them. This family
is parametrized by a parameter $p\in [0,1]$ such that:
\begin{enumerate}

\item For $p=0$ the algebra of observables has the one-dimensional
representation. The corresponding
states are one-component (scalar fields).

\item For $p\in(0,1)$, particularly for $p=\frac{1}{2}$,
the algebra of observables has the two-dimensional spinor
representation. The states
are $2$-spinors (spinor fields).

\item For $p=1$ the algebra of observables has a reducible
representation, which is the direct sum of
two one-dimensional representations. The corresponding states are
two-components ($2$-vector fields).
\end{enumerate}

One can consider the parameter $p$ as a value (without sign) of
possible projection of the spin of particles. Then the
two-projections quantization introduces in a natural way the existence
of three main types of relativistic particles (corresponding to zero,
one-half and unit spins). There is not an {\em elementary\/} particle with spin
higher that $1$ in our construction.

We make a very short (and a little bit skeptical) overview of the
standard relativistic quantizations (quantum field theory) in
Section~\ref{se:overview}. The background of our approach to this
problem (quantum and relativistic projections) will be introduced in
Section~\ref{se:projections}. In Section~\ref{se:models} we
consider two models of the two-projections relativistic quantization.
The first model is rather remote, nevertheless it demonstrates the
appearance of spin-like structure in our setting. The second model
describes a free relativistic particle with arbitrary spin in the
Minkowski space-time.

The problem of a joining of quantum theory and general covariance is
not a simple one. For example, in~\cite{ConnRov94} it was suggested to
make the notion of time even more relativistic than usual, i.e. the
time flow should depend even from states of systems at hands. At the
present paper another logical opportunity is studied: we admit a
violence of relativistic invariance on the quantum level, i.e. the
quantum and relativistic projections may not commute (see
Remark~\ref{re:noncommute}). Such an
assumption gives a description of observable spin effects.

There are some remarks on the paper style. The paper is addressed both
to physicists and mathematicians.
This explains why (sometimes) the explanation is too basic
for someone. The bibliography on the subject is enormous. We
usually refer only to the recent publication(s), which allow to
reconstruct a wider set of references.

It is a pleasure to express my thanks to Yu.~G.~Gurevich,
V.~V.~Krav\-chen\-ko, M.~V.~Kuzmin, B.~Melnik, Z.~Oziewicz,
I.~Spitkovsky, and B.~A.~Veytsman for useful discussions.

\section{Standard Quantum Field Theory and Relativistic Quantization}
\label{se:overview}

There are many (deeply intervening) ways to construct quantum field
theory. Two important streams are:
\begin{itemize}
\item The path integral techniques: the Wiener functional and the
Feynman
path integrals~\cite{FeynHibbs65,GlimmJaffe81}. We will discuss this
approach elsewhere~\cite{Kisil95f}.

\item The second quantization technique: the Dirac-Fock-Jordan-Wigner
approach~\cite{Dirac67}. \\
\end{itemize}
The last one is a two step construction:
\begin{enumerate}
\item One should construct a relativistic equation for a single
particle.
\item The second quantization procedure joins particles in the bosonic
and/or fermionic Fock spaces~\cite{Dirac67,Segal90}. This provides
a description of interactions, decays, creations, and annihilations of
particles.
\end{enumerate}

Herein we will focus on the first step---construction of relativistic
equations of a single relativistic particle.
The spin of a particle is the main characteristic of the
equation. Other qualities of particles (mass, charge, etc.) are only
parameters in the equation. But the type of equation itself sharply
depends on particle's spin. Moreover, the type of second quantization
is strictly  predestined by the type of statistic (Fermi-Dirac or
Bose-Einstein) and the  statistic in its turn is also determined by
the spin.

The spin is usually associated with representations of the
inhomogeneous Lorentz
group. Our consideration will show (Subsection~\ref{ss:quarel}) that
connection between the notion of spin and the theory of special
relativity is more
deeper than only a representation of the inhomogeneous Lorentz group.

The equation for a single particle is usually constructed as
Schr\"o\-din\-ger type equation
and the usual pre-requests\footnote{For the sake of simplicity all
notions are illustrated by their simplest (original) form.}  are:
\begin{itemize}
\item It should have a Lorentz invariant Hamiltonian. The fundamental
relativistic metric form
\begin{displaymath}
dt^2-dx_1^2-dx_2^2-dx_3^2
\end{displaymath}
is the most natural Lorentz invariant object and it generates the
wave equation (the Klein-Gordon equation without mass):
\begin{displaymath}
\left(\frac{\partial^2 }{\partial t^2 }-\frac{\partial^2 }{\partial
x_1^2 }-\frac{\partial^2 }{\partial x_2^2 }-\frac{\partial^2
}{\partial x_3^2 }\right) f(x)=0.
\end{displaymath}
If one considers the Einstein expression for the energy of a free relativistic
particle~\cite[(9,7)]{LandLifshII}
\begin{displaymath}
E=\sqrt{p^2c^2+m^2c^4}
\end{displaymath}
then the standard rules of quantization
\begin{displaymath}
E\rightarrow i\hbar\frac{\partial }{\partial t},\ \
p_i=-i\hbar\frac{\partial}{\partial x_i}
\end{displaymath}
give the {\em square-root Klein-Gordon\/} equation for a free
relativistic particle
\begin{equation}\label{eq:square-root}
 i\hbar\frac{\partial \phi(t)}{\partial t}=\sqrt{-
c^2\hbar^2\Delta+m^2c^4}\,\phi(t).
\end{equation}
Here $\Delta=\sum_1^3 \frac{\partial^2}{\partial x_j^2}$ is the Laplace
operator.
\item The equation under consideration should describe the number of
freedom degrees corresponding
to  the spin of particle.  The simple way for
construction~\eqref{eq:square-root} does not meet this request.

\end{itemize}

Meanwhile the first condition is a very natural one, it is my
impression, that the second condition was usually achieved by a
hand-made work (if not to say artificial). For
example~\cite[\S~4]{BogShir80}, for a free particle with spin $1$ it
is usually {\em a~priori\/} assumed that its states are described by a
$4$-vector field. Some weak motivation for this is the following:
$4$-vector field is a natural Lorentz invariant object. But
immediate application of the Klein-Gordon equation shows that such
states connected with negative energy solutions. To eliminate them one
applies additional constraints and obtains states, which are described
by $3$ independent components only.

\begin{figure}[t]
\begin{center}
{ \begin{tabular}{||c||c|c|c||}
\hline\hline
 Spin      & spin $0$      & \rule[-10pt]{0pt}{26pt}spin $\frac{1}{2}$
& spin $1$      \\ \hline
 Particle  & $\pi$-meson   &  electron          & photon        \\
\hline
Equation   & Klein-Gordon  & Dirac (factorized & Klein-Gordon   \\
           &               & Klein-Gordon)     &           \\ \hline
States     & scalar fields ($1$)& two-spinors fields ($4$) &
$4$-vectors fields  ($4$) \\ \hline
Correction & scalar fields ($1$)& spinors fields ($2$)  & $3$-vectors
fields ($3$)\\ \hline \hline
\end{tabular}}
\end{center}
\caption{Particles with different spins, their equations, types of
states and their dimensionality (shown in braces) by the standard
theory and results of correction by experiments.}\label{fi:particles}
\end{figure}

Another example is the Dirac equation for a free relativistic electron
(see~\cite{Dirac78}, \cite[\S~5]{BogShir80}
and~\cite[\S~1.1]{Thaller92}). It was
mentioned in Introduction, its deduction contradicts all usual rules
of quantization. Moreover, the Dirac ``factorization procedure'' was
never used to any other problem in physics and thus may hardly be
named a method\footnote{``What is the difference between method and
device? A method is a device, which you use
twice''~\cite[p.~208]{Polya57}.}.
Till now it was the only way to make the number of freedom degrees
large enough for a description of the spin of
the electron and its magnetic moment. But the Dirac equation provides
us with too many degrees of freedom and half of them correspond again
to negative energy solutions. This involves many
unpleasant consequences (Zitterbewegung, superluminal propagation,
etc.~\cite{Thaller92}). To eliminate them one should introduce, for
example, the Dirac ``holes in the sea of negative energy electron
states''.

We summarize information about different type of particle in
Figure~\ref{fi:particles}. We
would like to avoid the question: {\em which types of elementary
particles do exist, and which of them do are elementary\/}. At least
photon and electron are often believed to exist and be rather
elementary.

Our brief consideration justifies the following claim: {\em there is
no any unified and natural procedure to obtain relativistic equations
for different types of particles with right number of degrees of
freedom\/}.

We will present in the next Sections a procedure of relativistic
quantization based on the notion of casual structure. It is in an
agreement with the ``classical''
non-relativistic quantizations and gives a simple description for
the spin structure.
\begin{rem}
If the reader is familiar with the Einstein-Podolsky-Rosen par\-a\-dox
then the question arises: {\em May standard quantum mechanics be
combined with the notion of casualty at all\/}? I do not know the
answer to this question, but would like to make two observations (see also
Remark~\ref{re:noncommute}):
\begin{enumerate}
\item The Einstein-Podolsky-Rosen paradox is deeply connected with the
theory of measurements and interpretation of quantum mechanics. We do
not touch these topics herein.
\item The disagreement between the casualty structure and non-locality
of standard quantum mechanics is not the only contradiction in quantum
theory.
\end{enumerate}
\end{rem}

\section{Origins of Two Projections}\label{se:projections}

It this Section we explain how two projections arise in our approach.
Indeed, two words from the paper title---{\em relativistic\/} and {\em
quantization\/}---explain the existence of two projections.

\subsection{Non-Relativistic Quantizations Defined by
Projection}\label{ss:bargmann}

First, let us remind that the standard non-relativistic quantization
may be obtained by application of a projection to the classical
system under consideration. We give only a short summary of this
topic, the relevant information may be found in
\cite{Berezin74,Berezin75,BergCob87,Coburn90,Coburn94a,CobXia94,%
Guillemin84}
and their references.

Let $\FSpace{L}{2}(\Space{C}{n},d\mu_{n})$ be a space of all
square-integrable functions on
\Space{C}{n} with respect to the Gaussian measure
\begin{displaymath}
d\mu_{n}(z)=\pi^{-n}e^{-z\cdot\overline{z}}dv(z),
\end{displaymath}
where $dv(z)=dxdy$ is the usual Euclidean volume measure on
$\Space{C}{n}=\Space{R}{2n}$. The
Segal-Bargmann~\cite{Bargmann61,Segal60} (or the Fock) space
$\FSpace{F}{2}(\Space{C}{n})$ is the subspace of
$\FSpace{L}{2}(\Space{C}{n},d\mu_{n})$ consisting of all entire
functions, i.e. such functions $f(z)$ that
\begin{displaymath}
\frac{\partial f}{\partial \bar{z}_j}=0, 1\leq j \leq n.
\end{displaymath}
Denote
by $P_Q$ the orthogonal Bargmann projection \cite{Bargmann61}  of
$\FSpace{L}{2}(\Space{C}{n},d\mu_{n})$ onto the Segal-Bargmann (Fock) space
$\FSpace{F}{2}(\Space{C}{n})$. Then the formula
\begin{equation}
k(q,p)\rightarrow T_{k(q+ip)}=P_Q k(q+ip)I
\end{equation}
defines Berezin-Toeplitz (anti-Wick) quantization, which maps a
function
$k(q,p)=k(q+ip)$ on $\Space{R}{2n}=\Space{C}{n}$ to the Toeplitz operator $T_k$
 with the
pre-symbol $k(q+ip)$ on $\Space{C}{n}=\Space{R}{2n}$. There is an
identification
between the
Berezin quantization and the Weyl
quantization~\cite{Berezin74,Coburn90,Guillemin84}. The identification
has the simplest form for observables depending only on $p$ or $q$
alone. We will use it in Subsection~\ref{ss:momentum} for the
construction of the Schr\"odinger type equation of a free particle at
the Minkowski space.

\begin{example}~\cite{BergCob87} At the Segal-Bargmann
representation the operators of creation and annihilation of a
particle
at the $j$-th state are $a^+_j=z_jI$ and $a^-_j=\frac{\partial }{\partial z_j}$
correspondingly. Let us consider harmonic
oscillator with $n$ degrees of freedom. Its Hamilton function is
\begin{displaymath}
H(q,p)=\frac{1}{2}\sum_{j=1}^{n}(q_j^2 + p_j^2).
\end{displaymath}
Then the corresponding quantum Hamiltonian is the operator
\begin{displaymath}
T_{H(q,p)}= \frac{1}{2} P_Q \sum_{j=1}^{n}(q_j^2 +
p_j^2)I=nI+\sum_{j=1}^{n}z_j\frac{\partial }{\partial z_j}.
\end{displaymath}
\end{example}
\begin{rem}
The Berezin-Toeplitz  quantization is not the only quantization
generated by a projection. Let us remind that {\em geometric
quantization\/} procedure~\cite{Kirillov90,Woodhouse80} consists of
two steps: prequantization and quantization. The second step is,
in fact, restriction of operators achieved by prequantization to
manifold defined by polarization (projection to functions depending
only on ``coordinates'', roughly speaking).
\end{rem}

\subsection{Casual Projection in Relativistic Mechanics}

\begin{figure}
\begin{center}
\unitlength=1.2mm

\special{em:linewidth 0.4pt}
\linethickness{0.4pt}
\thicklines
\begin{picture}(107.00,56.00)
\qbezier(0.00,22.00)(49.00,27.00)(73.00,0.00)
\qbezier(73.00,0.00)(95.00,36.00)(107.00,35.00)
\qbezier(0.00,22.00)(18.00,46.00)(32.00,46.00)
\put(74.00,14.00){\makebox(0,0)[cc]{{$M$}}}
\put(48.00,35.00){\makebox(0,0)[cc]{{$X$}}}
\put(31.00,29.00){\makebox(0,0)[cc]{{$T_X M$}}}
\put(87.00,41.00){\makebox(0,0)[cc]{{$V_X$}}}
\qbezier(34.00,38.00)(53.00,41.00)(66.00,31.00)
\put(51.00,38.00){\vector(1,0){18.00}}
\put(51.00,38.20){\vector(1,0){18.00}}
\put(62.00,29.00){\makebox(0,0)[cc]{{$\gamma(\tau)$}}}
\put(71.00,37.00){\makebox(0,0)[cc]{{$\vec{u}$}}}
\put(70.00,24.00){\line(5,4){26.00}}
\put(30.00,44.80){\line(-5,-4){26.00}}
\put(96.00,44.80){\line(-1,0){66.00}}
\put(4.00,24.00){\line(1,0){66.00}}
\qbezier(107.00,35.00)(104.00,56.00)(32.00,46.00)
\put(51.03,38.00){\circle{0.8}}
\thinlines
\put(51.00,38.00){\line(6,1){37.00}}
\put(52.00,38.00){\line(6,-1){29.00}}
\put(57.00,39.00){\line(3,-2){3.50}}
\put(63.00,40.00){\line(3,-2){7.50}}
\put(69.00,41.00){\line(3,-2){11.50}}
\put(75.00,42.00){\line(3,-2){8.00}}
\put(81.00,43.00){\line(3,-2){3.50}}
\end{picture}
\end{center}
\caption[The casual structure for a dynamical system.]{
        \protect\parbox[t]{11.9cm}{The casual structure for a dynamic
system:\protect\\
                   \begin{tabular}[t]{c@{---}p{9.7cm}}
                   $M$&the space of events (space-time);\protect\\
                   $\gamma(\tau)$&a trajectory of the dynamical system
                      parametrized by the proper ``time''
$\tau$;\protect\\
                   $X$&a point of the trajectory $\gamma(\tau)$ for
                        the value $\tau=\tau_0$;\protect\\
                   $T_X M$&the tangent space at $X$;\protect\\
                   $V_X$&the casual cone in $T_X M$;\protect\\
                   $\vec{u}$&the velocity vector of the trajectory
                               $\gamma(\tau)$ at $X$ should
belong to
                               $V_X$.
                   \end{tabular}}}\label{fi:future}
\end{figure}

Now we introduce the second projection. Our consideration is based on
the book~\cite[Chap.~II]{Segal76}.  This is an alternate approach to
the theory of special relativity, which suggests that
variations from the Lorentz group may be useful. The main idea
is: one can introduce a general axiomatic relativistic structure not
by means of the Lorentz group, but by the usage of the notion of
casual structure. Namely, there is an ``infinitesimal future cone'' in
the tangent space at each point to the space of events (space-time).

The axiomatic formulation and mathematical implementation may be found
in~\cite[Chap.~II]{Segal76}. We give only a short illustration here
(see Figure~\ref{fi:future}). Let $M$ be the space of events
(space-time). We need not specify the dimensionality of $M$. Let
$\gamma(\tau)$ be a trajectory of the dynamical system (point) under
consideration parametrized by the proper ``time'' $\tau$. Let $X$ be a
point of $M$ lying on the trajectory $\gamma(\tau)$ for the value
$\tau=\tau_0$. We denote by $T_X M$ the tangent space at the point
$X$. The existence of the casual structure on $M$ implies that in
$T_X M$ there is the casual (or future, or light) cone, namely
$V_X$. Thus the velocity vector $\vec{u}$ of trajectory
$\gamma(\tau)$ at $X$ should
belong to the future cone  $V_X$.

The casual constraint may be achieved by the multiplication of the
Lagrangian function $L(q,\dot{q})$ by the characteristic function
$\chi_R(q,\dot{q})$ of the future cone. The new Lagrangian
$\chi_R(q,\dot{q})
L(q,\dot{q})$ is non-zero only for admissible points of tangent bundle
$T M$ and thus allows only relativistic motion. We will denote the
casual projection by $P_R=\chi_R I$. We would like to stress an
analogy between the transition from classical mechanics to quantum by
means of projection $P_Q$ and the passing from non-relativistic
mechanics to relativistic one by $P_R$.
\begin{example} The ``classical'' example is the
four-dimensional Minkowski space-time with the
pseudo-Euclidean metric (\cite[Chap.~IX]{Lanczos70}, \cite[Chap.~I,
II]{LandLifshII})
\begin{displaymath}
ds^2=c^2dx_0^2 -dx_1^2-dx_2^2-dx_3^2.
\end{displaymath}
The casual structure is defined by the future part of the light cone:
\begin{displaymath}
c^2dx_0 \geq \sqrt[]{dx_1^2+dx_2^2+dx_3^2}.
\end{displaymath}
The relativistic projection $P_R$ here is the operator of
multiplication by the characteristic function $\chi_R(p)$ of the
future cone. Note, that in this case the function $\chi_R$ does not
depend on $q$ and this will greatly simplify our consideration in
Subsection~\ref{ss:momentum}.

Usually textbooks link the theory of special relativity with the
Lorentz
invariance of the theory. We will not touch the group of relativistic
transformations herein. However, our theory (the relativistic
projection $P_R$ and the Hamiltonian function) will be defined purely
in terms of the light cone (see Section~\ref{se:models}). Thus our
theory will be invariant under all relativistic transformations, which
(by their definition) preserve the light cone.
\end{example}
\begin{rem}
It is interesting, that in classical mechanics we do not need
such relativistic projection. Outside the light cone the Lagrangian
of a free particle~\cite[(95.8)]{Lanczos70}
\begin{displaymath}
L=-c^2m\sqrt{1-\frac{1}{c^2}(\dot{q}^2_1+\dot{q}^2_2+\dot{q}^2_3)}
\end{displaymath}
is purely imaginary and thus is out classical theory. In quantum theory the
complex numbers are on an equal footing with the real ones and
separation of permitted and prohibited parts of phase space should be
done explicitly.
\end{rem}
\begin{rem} \label{re:anti} We have considered only the {\em future\/}
part of the light cone. But it also has the {\em past\/} part. Let us
remind that {\em anti\/}-particles may be considered as corresponding
{\em particles\/} moving backward in time ($CPT$
invariance~\cite[\S~13]{BerLif82}). To describe this one may
wish to multiply the Lagrangian function by the characteristic
function of the  past part of the light cone. In both cases (the
future and past part of light cone) (anti-)particles are moving
the forward in proper time. Thus particles and corresponding
anti-particles have appeared in our consideration on the equal
symmetrical footing.
\end{rem}

\section{Models of the Two-Projections Relativistic Quantization}
\label{se:models}

Previous consideration shows that a natural algebra of observables
for a quantum relativistic particle should be generated by (at least)
three operators: the quantum projection $P_Q$, the relativistic
projection $P_R$, and by the Hamilton function $H(q,p)$. We will
denote such \Cstar-algebra of observables by
$\algebra{O}(P_Q,P_R,H(q,p))$. This Section
is devoted to two concrete realizations of this algebra.

\subsection{Relativistic Quantization: a Toy Model}\label{ss:quarel}

First we will consider unrealistic case of the Hamiltonian function
$H(q,p)$ identically equal to $1$. Then algebra
$\algebra{O}(P_Q,P_R,H(q,p))$ is generated only by two projections
$P_Q$ and $P_R$:
\begin{displaymath}
\algebra{O}(P_Q,P_R,H(q,p))=\algebra{O}(P_Q,P_R)
\end{displaymath}

The algebra generated by two projections is very well studded in
mathematics (see, for example,~\cite{Spitkovsky94a,VasSpi81})
and have already appeared in quantum mechanics (``two questions
generate infinitely many questions''~\cite[Appendix~3]{Meyer93}). The
following result is the basis of our construction.

\begin{thm}~\textup{\cite{VasSpi81}} Let the points $0$ and $1$ be
non-isolated points of the spectrum $\sigma=\object{sp}(P_Q-P_R)^{2}$.
Then the algebra $\algebra{O}(P_{Q},P_R)$ generated by two
projections $P_{Q}$ and $P_{R}$ is
isometrically isomorphic to the algebra of all $2\times
2$ continuous matrix-functions on
$\sigma$, which are diagonal at the points of
$\{0,1\}\subset\sigma$. This isomorphism
$\phi$ is defined by the following mapping of the generators of
$\algebra{O}(P_{Q},P_R)$
\begin{eqnarray*}
\phi: P_{Q}&\longmapsto&
\left(\begin{array}{cc}
1-p           & \sqrt {p(1-p)}\\
\sqrt{p(1-p)} & p
\end{array}\right),\\
\phi: P_{R}&\longmapsto&\left(\begin{array}{cc}1& 0\\  0& 0
\end{array}\right),
\end{eqnarray*}
where $p \in \sigma$.
\end{thm}

The parameter $p$ appeared at the previous Theorem has the following
geometric meaning. The spectrum $\sigma$ characterize the mutual
disposition of
$\object{Im} P_Q$ and $\object{Im} P_R$ and, in some sense,
generalizes square of sine of the angle between two lines in the
two-dimensional case.
This geometric interpretation suggests to understand the parameter $p$
as a possible value of projection (without sign) of spin of the
particle under consideration.
\begin{rem}
In our approach values of projection of spin are not quantized in
the quantitative sense and may fill whole interval $[0,1]$. But they
do are ``quantized'' in the qualitative (phenomenon) sense. For $p=0$
and $p=1$ matrixes are diagonal (correspond to ``vector''
representations) and for $p\in (0,1)$ (particular $\frac{1}{2}$) they
are general matrixes (correspond to ``spinor'' representations). Thus
we will count all value $p\in (0,1)$ as corresponding to spin
$\frac{1}{2}$.
Probably, ``hidden internal degree of freedom'' $p$ may be employed in
future.
\end{rem}

Let us assume, that $\sigma=[0,1]$. We introduce the subalgebra
(even an ideal) $\algebra{O}'(P_Q,P_R)$ of the algebra
$\algebra{O}(P_Q,P_R)$ such that
\begin{displaymath}
\algebra{O}'(P_Q,P_R)=\{A\in \algebra{O}(P_Q,P_R) \such
Af=0 \mbox{ for all } f\in (\object{Im }P_Q)^{\perp} \cap (\object{Im
}P_R)^{\perp}\}.
\end{displaymath}
By other words, a non-quantum {\em and\/} non-relativistic behavior in
$\algebra{O}'(P_Q,P_R)$ is unobservable.
Then one obtains an evident
\begin{cor}\label{co:dimensionality}
For parameter $p\in [0,1]$  the algebra $\algebra{O}'(P_Q,P_R)$
has:
\begin{enumerate}

\item For $p=0$---the one-dimensional representation. The
corresponding
states are one-component (scalar fields);

\item For $p\in(0,1)$
(particularly for $p=\frac{1}{2}$)---the two-dimensional spinor
representation. The states are $2$-spinors (spinor fields).

\item For $p=1$--- a reducible representation, which is the direct sum
of
two one-dimensional representations. The corresponding states are
two-components ($2$-vector fields);
\end{enumerate}
\end{cor}
\begin{example}
For the observable $P_Q+P_R\in\algebra{O}'(P_Q,P_R)$ we have:
\begin{displaymath}
\begin{array}{ccc}
p=0      &        p=\frac{1}{2} \vspace{5pt}      &    p=1 \\
\left(\begin{array}{cc}
2&0\\0&0
\end{array}\right) &
\left(\begin{array}{cc}
\frac{3}{2}&\frac{1}{2}\vspace{5pt}\\ \frac{1}{2}&\frac{1}{2}
\end{array}\right)
&
\left(\begin{array}{cc}
1&0\\0&1
\end{array}\right)
\end{array}
\end{displaymath}
\end{example}

In our framework {\em elementary\/} particles with spin $0$, $\frac{1}{2}$,
and $1$ are only allowed. Particles with higher spin may be considered
as composite ones.

There are some conclusions from such interpretation:
\begin{itemize}
\item {\em Particle of a spin $s$ has $2s+1$ degree of freedom\/}. For
spin $0$, $\frac{1}{2}$, and $1$ the possible values of its projection
(without sign)
are $\{0\}$, $\{\frac{1}{2}\}$, and $\{0,1\}$ correspondingly.
Counting dimensionality of matching representations from
Corollary~\ref{co:dimensionality} one can obtain the assertion.

\item {\em If a particle has the luminal propagation it should
have the spin with projection $1$\/}. Indeed, for a particle with
the luminal propagation the relativistic projections $P_R$ is the
operator of multiplication by the characteristic function of the
boundary of light cone. But the boundary of a reasonable cone has the
zero measure, thus $P_R=0$. From here
\begin{displaymath}
\object{sp}(P_Q-P_R)^2=\object{sp}P_Q^2=\object{sp}P_Q=\{0,1\}.
\end{displaymath}
We should exclude the value $0$, because it corresponds to a
non{-}quan\-tum and non{-}relativistic behavior. After that the only
possibility is the projection with value $1$. Note, that our
conclusion is in the total agreement with the case of photons.

\item {\em If a particle has an underluminal propagation then it
may have the spin $\frac{1}{2}$\/}. For a particle with underluminal
propagation the relativistic projection $P_R$ is not zero and
$\object{sp}(P_Q-P_R)^2$ may contain more points than $0$ and $1$. It
will depend on additional constrains, which an opportunity for the
projection will realize.
\end{itemize}
\begin{rem}\label{re:noncommute}
In the given model particles with the spin $0$ or $1$ are defined by commuting
projections $P_Q$ and $P_R$, i.e. their quantum and relativistic natures are in
the agreement. In contrary, the spin $\frac{1}{2}$ may arise only for
non-commuting projections
$P_Q$ and $P_R$. Thus {\em an existence of particles with spin $\frac{1}{2}$
may be explained by a non-compatibility of the relativity with the quantum
world\/}.
\end{rem}
It seems that first conclusions even from the very toy model are
natural, thus we are going to construct more realistic model.

\subsection{Realistic Hamiltonian and Improved Equation for a
Free Relativistic Particle}

We have seen in the previous Subsection that a spin-like structure has
arisen in our approach from the very existence of two projections and
does not depend on the properties of Hamiltonian function. Thus for
particles with all types of spin we can select a {\em scalar\/}
Hamiltonian function guiding by relativistic non-quantum mechanics.

Let us consider the Minkowski four-dimensional space-time (space of
events)
with the pseudo-Euclidean metric (\cite[Chap.~IX]{Lanczos70}
and~\cite[Chap.~I, II]{LandLifshII})

\begin{displaymath}
ds^2=c^2dx_0^2 -dx_1^2-dx_2^2-dx_3^2.
\end{displaymath}
The casual structure is defined by the future part of the light cone:
\begin{equation}\label{eq:light}
c^2dx_0 \geq \sqrt[]{dx_1^2+dx_2^2+dx_3^2}.
\end{equation}
Here the relativistic projection $P_R$ is the operator of
multiplication by the characteristic function $\chi_R(p)$ of the
future cone (for anti-particles see Remark~\ref{re:anti}). The quantum
projection $P_Q$ is the Bargmann projection
on the four-dimensional Segal-Bargmann space
$\FSpace{F}{2}(\Space{C}{4})$. It was already
calculated~\cite{KiRaTruVa94a} that
\begin{lem}
The spectrum of
the operator
\begin{displaymath}
(P_{Q}-P_R)^2|_{F_2(\Space{C}{n})}
\end{displaymath}
acting on $F_{2}(\Space{C}{n})$ is equal to $[0,1]$.
\end{lem}
Thus in this model particles of spin $0$, $\frac{1}{2}$, and $1$ are
all permitted.

Let us find the Hamilton function. The variation of action integral
for a free particle with mass $m$ (\cite[(95.8)]{Lanczos70},
\cite[(8,1)]{LandLifshII})
\begin{displaymath}
S=-mc\int_a^b ds
\end{displaymath}
gives equation~\cite[(9,10)]{LandLifshII}
\begin{displaymath}
\frac{d u_j}{d s}=0, \mbox{ where } u_j=\frac{d x_j}{ds}.
\end{displaymath}
The Hamilton formulation may be achieved by introduction of
$4$-momentums
(\cite[(95.15)]{Lanczos70}, \cite[(9,14)]{LandLifshII})
\begin{displaymath}
p_j=mc \frac{dx_j}{ds} = mc u_j
\end{displaymath}
and Hamilton function of a free particle (\cite[96.8]{Lanczos70})
\begin{equation}\label{eq:hamiltonian}
H(q,p)=\frac{1}{2mc}\sum_{j=0}^3 p_i p^i= \frac{1}{2mc}(p_0^2-p_1^2-
p_2^2-p_3^2).
\end{equation}
The Hamiltonian~\eqref{eq:hamiltonian} is manifestly Lorentz invariant
and positive inside the light cone.

Thus the algebra $\algebra{O}'(P_Q,P_R,H(q,p))$ of observables of a
free particle in the
Segal-Bargmann representation is the algebra generated by quantization
projection $P_Q$ (Subsection~\ref{ss:bargmann}), relativistic
projection $P_R$ of multiplication by the characteristic function of
the future cone~\eqref{eq:light} and the Hamiltonian
$H(q,p)$~\eqref{eq:hamiltonian}. The Heisenberg equation for an
observable $A\in \algebra{O}'(P_Q,P_R,H(q,p))$ may be set as follows
\begin{displaymath}
\frac{d A(\tau)}{d\tau}=\frac{i}{\hbar}[A(\tau),H_Q],
\end{displaymath}
where $H_Q=P_Q P_R H(q,p)I$ is the quantum (operator) Hamiltonian and
$\tau$ is the proper time in the Minkowski space.

The corresponding Schr\"odinger equation for a state $\phi(\tau)$
(which has a spin structure by the existence of two projections) has
the form
\begin{equation}\label{eq:schrodinger}
\frac{\partial \phi(\tau)}{\partial \tau}=-\frac{i}{\hbar} H_Q
\phi(\tau).
\end{equation}

\subsection{The Relativistic Equation in the Schr\"odinger
Representation}\label{ss:momentum}

Now we would like to write equation~\eqref{eq:schrodinger} in the
Schr\"odinger representation. To this end we need transfer operator
$H_Q=P_Q P_R H(q,p) I$ from the Segal-Bargmann representation to the
Schr\"odinger one. It is easy to do because function $P_R H(q,p)$ for
a free particle in a flat space-time depends on variables $p$ only. We
produce our calculations in the momentum representation.

We will use the following standard notations:
\begin{displaymath}
z=x+iy=(z_{1},\ldots,z_{n})\in\Space{C}{n}.
\end{displaymath}
Let $\overline{z}=(\overline{z}_{1},\ldots,\overline{z}_n)$
with the usual notion of the complex conjugation. For $z, w \in
\Space{C}{n}$ let
\begin{eqnarray*}
z\cdot w &=&z_{1}w_{1}+\ldots+z_{n}w_{n},\\
|z|^{2}  &=&|z_{1}|^{2}+\ldots+|z_{n}|^{2}(=z\cdot\overline{z}),\\
(x,y)&=&(x_{1},\ldots,x_{n},y_{1},\ldots,y_{n})\in
\Space{R}{2n}=\Space{R}{n}\oplus\Space{R}{n}.
\end{eqnarray*}
Denote by $d\mu_{n}(z)$ the following Gaussian measure over
$\Space{C}{n}$%
$$d\mu_{n}(z)=\pi^{-n}e^{-z\cdot\overline{z}}dv(z),$$
where $dv(z)=dxdy$ is the usual Euclidean volume measure on
$\Space{C}{n}=\Space{R}{2n}$.

To do calculations let us introduce the following
operators (see~\cite{KiRaTruVa94a} for details of calculations).
Introduce the unitary operator
$$U:L_{2}(\Space{C}{n},d\mu_{n})\rightarrow
L_{2}(\Space{R}{2n})=L_{2}(\Space{R}{2n},dxdy),$$
defined by
$$(U\varphi)(z)=\pi^{-{ n\over  2}}e^{-{ z\cdot\overline{z}\over 2
}}\varphi(z),$$
or
$$(U\varphi)(x,y)=\pi^{-{ n\over  2}}e^{-{ x^2+y^2\over 2
}}\varphi(x+iy).$$
The unitary operator $I\otimes F$, where
$$(F f)(y)=(2\pi)^{-{ n\over 2 }}\int_{\Space{R}n}e^{-i\eta\cdot
y}f(\eta)\,d\eta$$
is the Fourier transformation, maps isometrically the space
$$L_{2}(\Space{R}{2n},dxdy)=L_{2}(\Space{R}{n},dx)\otimes
L_{2}(\Space{R}{n},dy)$$
onto itself.
Now introduce the isomorphism
$$W=W^{*}=W^{-1}:L_{2}(\Space{R}{2n})\rightarrow
L_{2}(\Space{R}{2n}),$$
where
$$(Wf)(x,y)=f({ 1\over \sqrt {2}}(x+y),{ 1\over \sqrt {2}}(x-y)).$$
Introduce the isometrical imbedding
$$R_{0}:L_{2}(\Space{R}{n},dx)\rightarrow
L_{2}(\Space{R}{n},dx)\otimes L_{2}(\Space{R}{n},dy)$$
defined by
$$R_{0}:g(x)\mapsto g(x)\cdot l(y).$$
 Then the adjoint operator
$$R^{*}_{0}: L_{2}(\Space{R}{2n})\rightarrow  L_{2}(\Space{R}{n})$$
is defined by
$$(R^{*}_{0}f)(x)=\pi^{-{ n\over 4 }}\int_{\Space{R}n}f(x,y)\,e^{-{
1\over 2 }y^2}\,dy.$$
Now, the operator $R=R^{*}_{0} W(I\otimes F) U$ maps the space
$L_{2}(\Space{C}{n},d\mu_{n})$ onto
$L_{2}(\Space{R}{n},dx)$ and the restriction
$$R|_{F_2(\Space{C}n)}: F_{2}(\Space{C}{n})\rightarrow
L_{2}(\Space{R}{n},dx)$$
is an isometrical isomorphism.

The adjoint operator
$$R^{*}=U^{-1}(I\otimes F^{-1})W
R_{0}:L_{2}(\Space{R}{n},dx)\rightarrow F_{2}(\Space{C}{n})\subset
L_{2}(\Space{C}{n},d\mu_{n})$$
maps isomorphically and isometrically the space
$L_{2}(\Space{R}{n},dx)$ onto the Segal-Bargmann
space $F_{2}(\Space{C}{n})$.
We have the following representations
\begin{eqnarray*}
P_Q&=&R^{*}R:L_{2}(\Space{C}{n},d\mu_{n})\rightarrow
F_{2}(\Space{C}{n}),\\
I&=&RR^{*}:L_{2}(\Space{R}{n})\rightarrow L_{2}(\Space{R}{n}).
\end{eqnarray*}

Now we calculate the image
\begin{displaymath}
H_S=R H_QR^{*}=R P_Q P_R H(q,p)R^{*}
\end{displaymath}
of operator $H_Q$ under isometry $R$ between the Segal-Bargmann and
the Schr\"odinger representations.
\begin{eqnarray*}
H_S&=&R P_Q P_R H(q,p) R^{*}=\\
&=&R R^{*}\cdot R P_R H(q,p) R^{*}=R\chi_R(p) H(q,p) R^{*}=\\
&=&R^{*}_{0}W(I\otimes F)U\chi_R(p) H(q,p) U^{-1}(I\otimes F^{-
1})WR_{0}=\\
&=&R^{*}_{0}W(I\otimes F)\chi_R(p) H(q,p) I\otimes F^{-1})WR_{0}=\\
&=&R^{*}_{0}W\chi_R(p) H(q,p) R_{0}=\\
&=&R^{*}_{0}\chi_R({ \xi+x\over \sqrt {2 }}) H(q,{ \xi+x\over \sqrt
{2 }})  R_{0}.
\end{eqnarray*}
Here we use that $H(q,p)=H(p)= \frac{1}{2mc}\sum_{j=0}^3 p_i p^i$ does
not
depend on $q$.
Thus
\begin{eqnarray*}
(H_S f)(\xi)&=&\pi^{-1}\int_{\Space{R}{4}}e^{-{ 1\over 2
}x^2}\chi_R
({ \xi+x\over  \sqrt {2 }})H(\frac{\xi+x}{\sqrt[]{2}}) f(\xi) \pi^{-1} e^{-{
1\over 2 }x^2}\,dx=\\
&=&\lambda(\xi)f(\xi),
\end{eqnarray*}
where
\begin{eqnarray}
\lambda(\xi)&=&\pi^{-2}\int_{\Space{R}{4}}\chi_R({
\xi+x\over \sqrt {2 } })H(\frac{\xi+x}{\sqrt[]{2}})e^{-
x^2}\,dx \nonumber \\
&=&\frac{1}{4mc\pi^{2}}\int_{V-\xi} \sum_{j=0}^3(\xi+x)_j(\xi+x)^j\,
e^{-x^2}\,dx.\label{eq:lambda}
\end{eqnarray}

\begin{thm}
The Hamilton operator $H_S$ of a free relativistic particle in the
momentum Schr\"odinger representation is the unbounded operator
$H_S=\lambda(\xi)I$ of multiplication by the positive valued function
$\lambda(\xi)$~\eqref{eq:lambda}. In the Schr\"odinger coordinate
representation this operator is $H_S= F^{-1}\lambda(\xi)F$.

Operator $H_S$ have a positive spectrum $(0,+\infty)$ and is
Lorentz invariant.
\end{thm}

The Schr\"odinger equation in the coordinate representation takes the
form:
\begin{equation}\label{eq:coordinate}
\frac{\partial f(\tau)}{\partial \tau}=-\frac{i}{\hbar}
F^{-1} \lambda(\xi) F f(\tau).
\end{equation}

We would like to give an insight on the function $\lambda(\xi)$
because it is definitely not elementary one.

\begin{lem}
One has decomposition
\begin{displaymath}
\lambda(\xi)= \frac{1}{4mc} \chi_R(\xi) \sum_{j=0}^3 \xi_j \xi^j +
o(\sum_{j=0}^3 \xi_j \xi^j).
\end{displaymath}
\end{lem}
\begin{proof}{}
Let us remind that $\chi_R(\xi)H(\xi)$ is only of polynomial growth at infinity
on \Space{R}{4} and on the space of such function
\begin{displaymath}
\frac{t^4}{\pi^2} e^{-t^2y^2}\rightarrow \delta (y), \mbox{ where }
t\rightarrow \infty
\end{displaymath}
(in sense of generalized functions). Thus we have
\begin{eqnarray*}
\lim_{t\rightarrow\infty}
t^{-2}\lambda(t\xi)&=&\lim_{t\rightarrow\infty}\frac{1}{4mc\pi^{2}
t^{-2}}\int_{\Space{R}{4} } \chi_R({
t\xi+x})\sum_{j=0}^3(t\xi+x)_j(t\xi+x)^j\,
e^{-x^2}\,dx\\
 &=&\lim_{t\rightarrow\infty}\frac{1}{4mc\pi^{2} t^{-2}}\int_{\Space{R}{4} }
\chi_R({
\xi+\frac{x}{t}})t^2\sum_{j=0}^3(\xi+\frac{x}{t})_j(\xi+\frac{x}{t})^j
e^{-x^2} dx\\
 &=&\lim_{t\rightarrow\infty}\frac{1}{4mc}\int_{\Space{R}{4} } \chi_R({
\xi+\frac{x}{t}}) \sum_{j=0}^3(\xi+\frac{x}{t})_j(\xi+\frac{x}{t})^j
\frac{t^4}{\pi^{2}}e^{-t^2 (x/t)^2} d\frac{x}{t}\\
 &=&\lim_{t\rightarrow\infty}\frac{1}{4mc}\int_{\Space{R}{4} } \chi_R(
\xi+y) \sum_{j=0}^3(\xi+y)_j(\xi+y)^j
\frac{t^4}{\pi^{2}}e^{-t^2 y^2} dy\\
 &=&\lim_{t\rightarrow\infty}\frac{1}{4mc}\int_{\Space{R}{4} } \chi_R(
\xi+y) \sum_{j=0}^3(\xi+y)_j(\xi+y)^j
\left(\frac{t^4}{\pi^{2}}e^{-t^2 y^2} \right)dy\\
 &=&\frac{1}{4mc}\int_{\Space{R}{4} } \chi_R(
\xi+y) \sum_{j=0}^3(\xi+y)_j(\xi+y)^j\,
\delta(y)\,dy\\
&=&\frac{1}{4mc} \chi_R(\xi) \sum_{j=0}^3\xi_j \xi^j.
\end{eqnarray*}
Granting homogeneity of the Hamiltonian one obtains the assertion.
\end{proof}
By this decomposition one can see, that the equation
\begin{displaymath}
\frac{\partial f(\tau)}{\partial \tau}=-\frac{i}{4\hbar mc}
\left( \frac{\partial^2 }{\partial x_0^2} -\frac{\partial^2 }{\partial
x_1^2} -\frac{\partial^2 }{\partial x_2^2} -\frac{\partial^2
}{\partial x_3^2} \right) f(\tau)
\end{displaymath}
may be considered as the first approximation to
equation~\eqref{eq:coordinate}.

\begin{rem}
There are no principal difficulties to develop our formalism also for
a particle in the external field (i.e. the Hamiltonian explicitly
depending on coordinates $x_i$) and/or in the curved space-time (i.e.
the characteristic function of the future cone depending on
coordinates). But in this case one cannot expect the simplicity of
equation~\eqref{eq:coordinate}.
\end{rem}

\newcommand{\noopsort}[1]{} \newcommand{\printfirst}[2]{#1}
  \newcommand{\singleletter}[1]{#1} \newcommand{\switchargs}[2]{#2#1}
  \newcommand{\irm}{\mbox{\rm I}} \newcommand{\iirm}{\mbox{\rm II}}
  \newcommand{\vrm}{\mbox{\rm V}}

\end{document}